\documentclass[%
 reprint, 
 amsmath,amssymb,
 aps, 
prl,
]{revtex4-2}

\usepackage{graphicx}
\usepackage{dcolumn}
\usepackage{bm}

\begin{document}

\preprint{APS/123-QED}

\title{\textbf{Phase Matching of Orbital Angular Momentum in Rare Earth Ion Doped Solid State Systems} 
}%

\author{Owen R. Wolfe}
     \email{Contact author: owen.wolfe@montana.edu}
\author{Joshua Dugre}%
\author{Grant Kirkland}
\author{R. Krishna Mohan}
\affiliation{%
 Spectrum Lab, Montana State University, Bozeman, Montana.
}%

\date{\today}

\begin{abstract}
In this work, we demonstrate the generation of stimulated photon echos carrying unique topological charge that results from the temporal phase matching conditions of three independent beams spatially and spectrally overlapped in a cryogenically cooled rare earth ion doped solid state system. A sample of Tm\(^{3+}\):YAG was used to generalize the momentum phase matching condition to include the orbital angular momentum of the input fields. The input fields and corresponding photon echo were characterized via astigmatic transform, with results mapping directly to the expected behavior of traditional stimulated photon echos. These results demonstrate that rare earth ion doped systems are capable of spatial multiplexing, spatial filtering, and the generation of structured light, opening pathways towards real time optical multi-modal signal processing.
\end{abstract}

\maketitle

\begin{figure}[h]
\includegraphics[scale=.2]{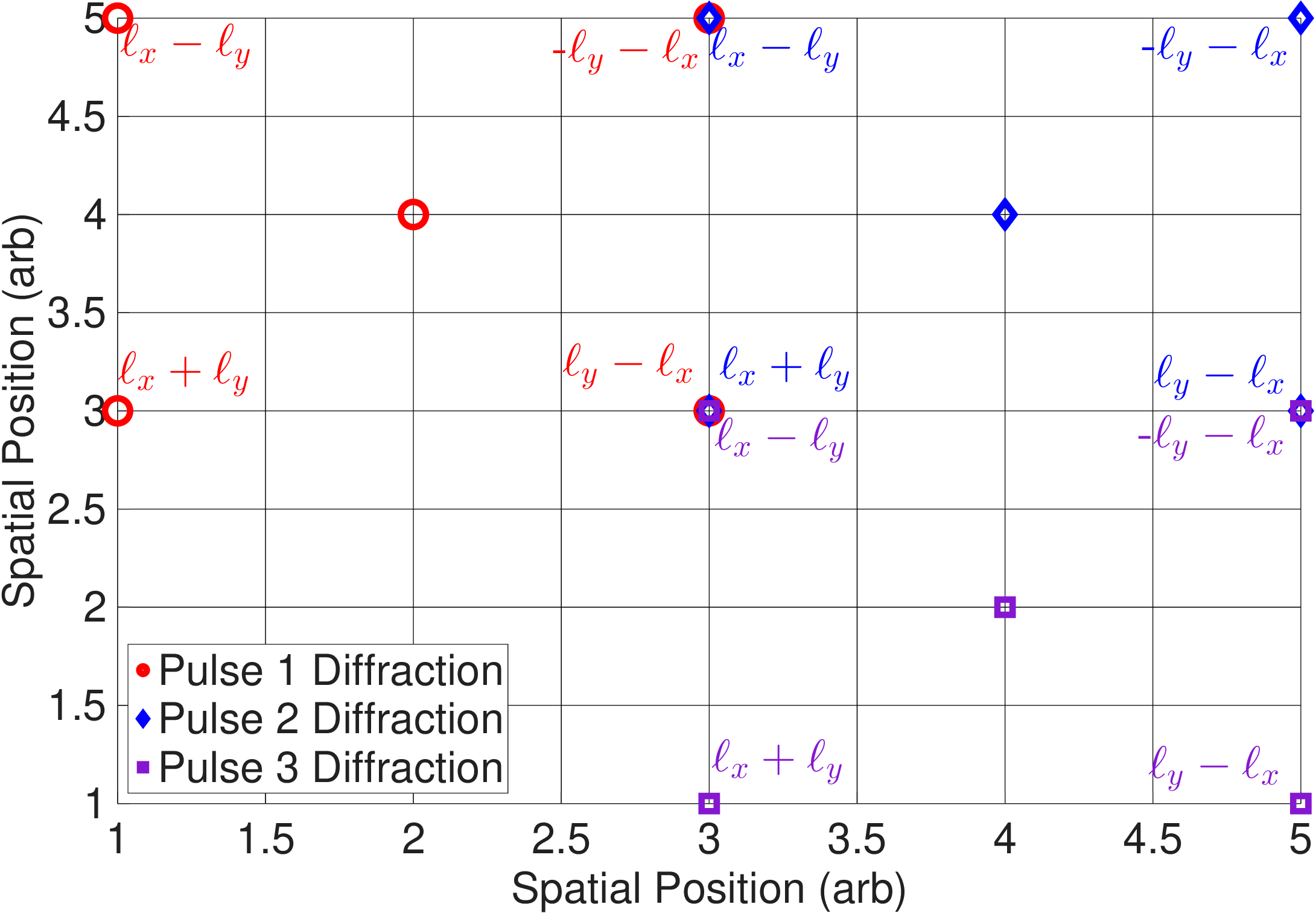}\caption{\label{fig:DVG_Schematic} The DVG will produce a strong zero order mode as well as four diffracted orders for each input beam, each carrying a unique topological charge $\ell$. Given the initial input geometry, the equal spacing of the imaged beams will result in a single spatially overlapped beam, each carrying a different $\ell$ (represented at position [3,3]). This beam is then spatially isolated and imaged into the REI material for processing.}
\end{figure}

\begin{figure*}
\includegraphics[width=.8\textwidth]{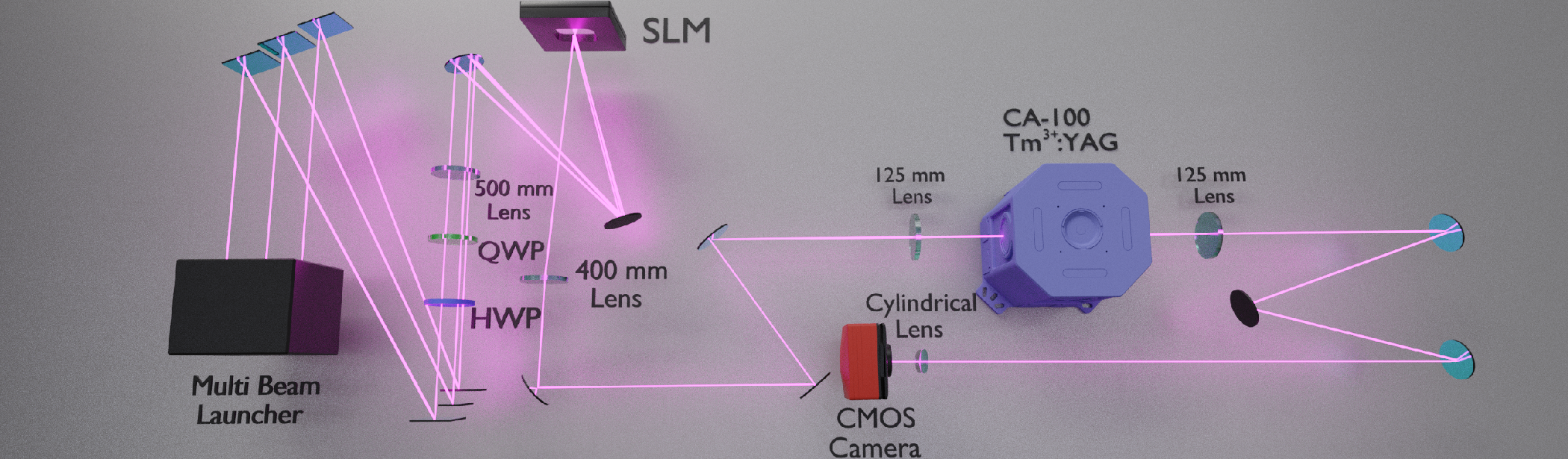}
\caption{\label{fig:OAM_Echo_Schematic} A schematic of the free space optical system used to generate and detect SPEs with unique OAM order. All three optical inputs were first polarized in the preferential plane of the SLM before being imaged onto the center of the phase screen to generate three overlapping Laguerre-Gauss beams. These beams were then collimated before being imaged into the sample of Tm\(^{3+}\):YAG contained within the CA-100. The resulting output was then collimated before being directed onto a CMOS camera for image analysis, with the camera located at the focal plane of a cylindrical lens when performing astigmatic transforms to calculate the topological charge. To isolate the SPE, the camera was triggered using the same multi-channel pulse generator that controlled the programming sequence. Definitions: HWP, Half wave plate; QWP, Quarter wave plate; SLM, spatial light modulator; CA-100, CryoAdvance 100.}
\end{figure*}

\begin{figure*}
\includegraphics[scale=0.45]{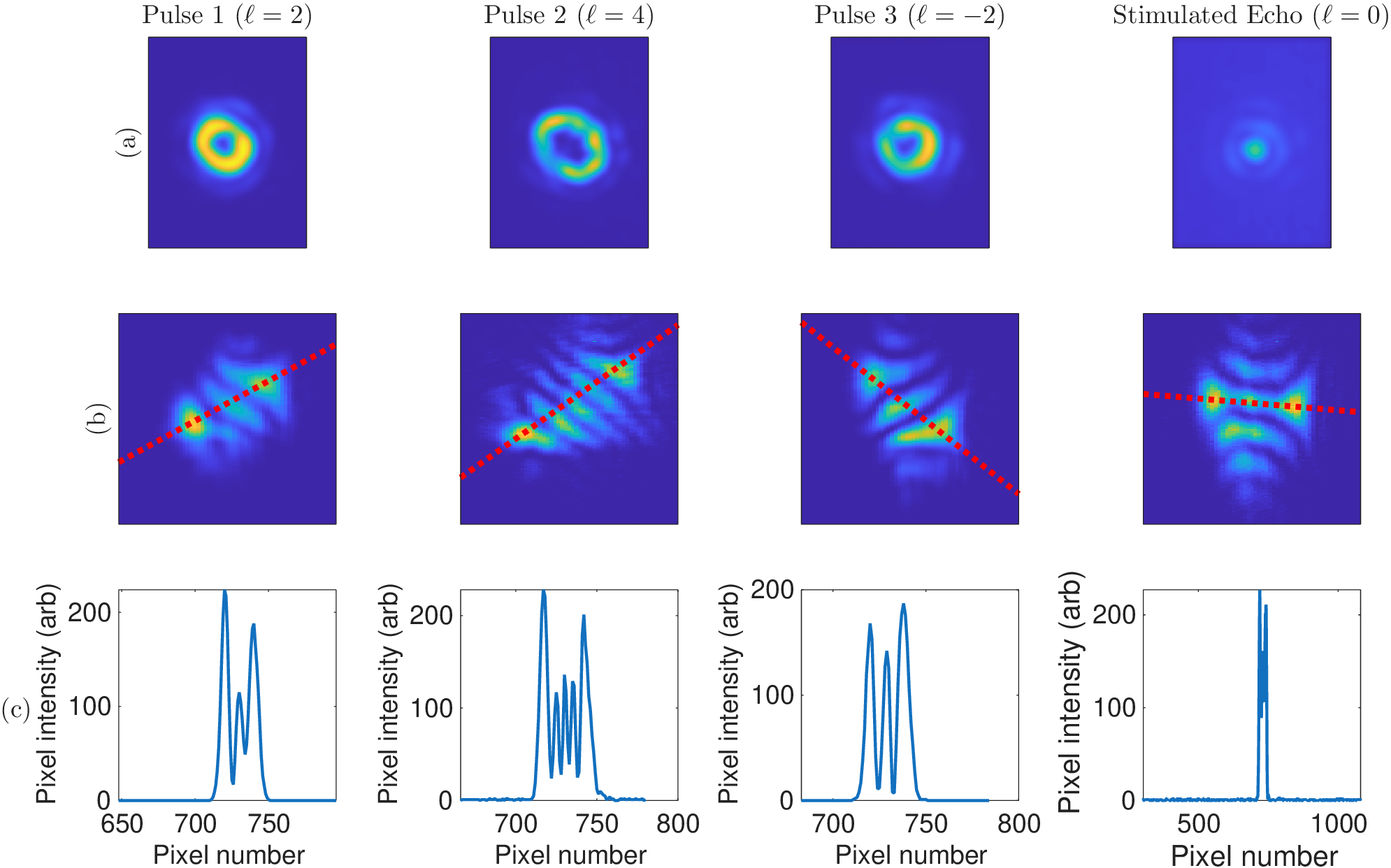}
\caption{\label{fig:OAM_Images} a.) Captured images of the modes of pulses 1, 2 and 3 as well as the SPE. Each of the pulses shows a clear central null indicative of a mode which carries OAM. The stimulated echo however shows no central null and has an Airy disk like intensity distribution. b.) Captured images of the intensity profile of each pulse and the stimulated echo at the focal plane of the cylindrical lens. After the astigmatic transform applied by the cylindrical lens, the number of nulls in each image indicates the topological charge of the incident beam. C.) A 1-D slice of each image taken along the doted line shown in the astigmatic transform images.}
\end{figure*}

\begin{figure}[h]
\includegraphics[scale=.125]{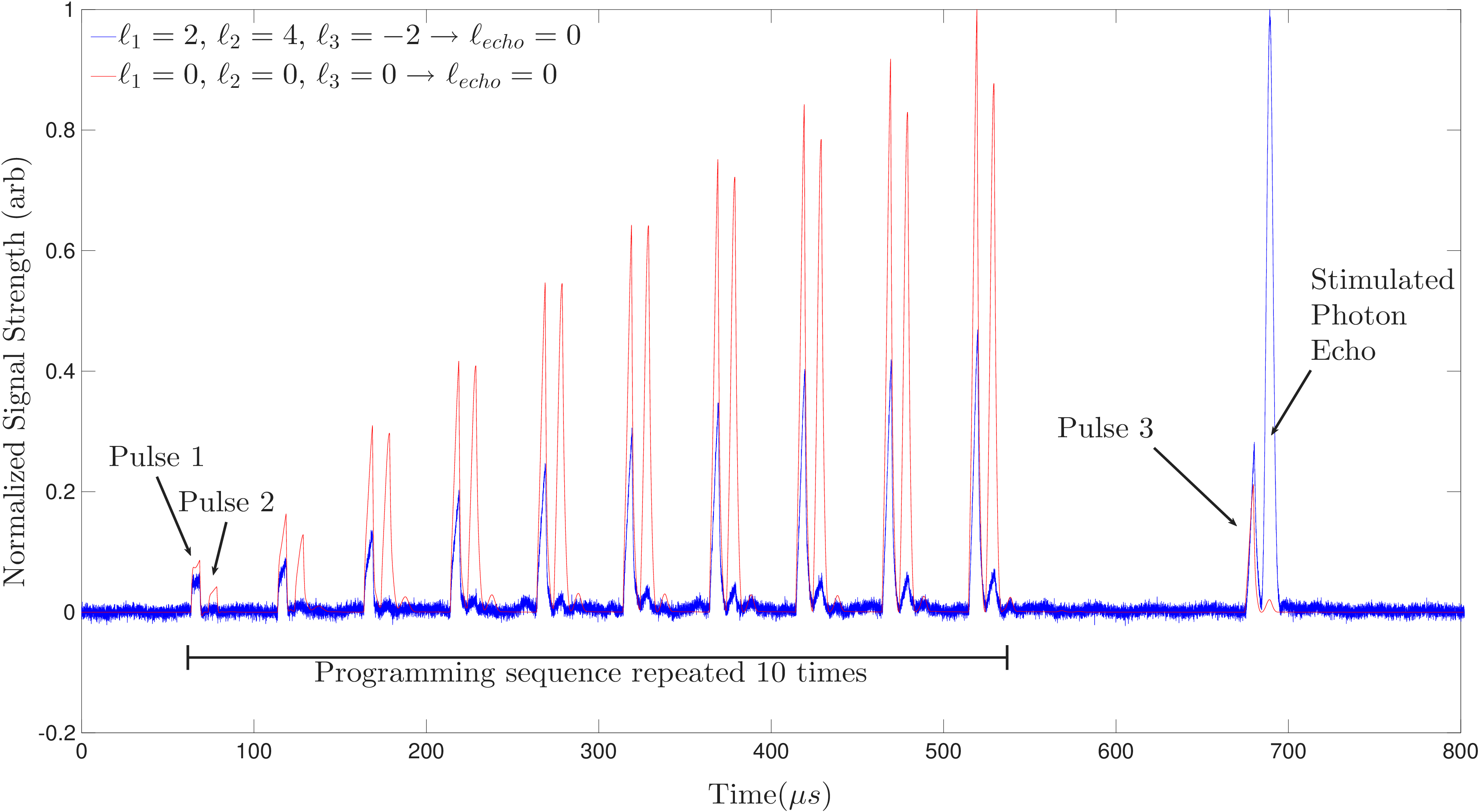}\caption{\label{fig:time_domain_echoes} Time domain traces showing the mode filtered echo  for $\ell_1 = 0$, $\ell_2 = 0$, $\ell_3 = 0$ as well as $\ell_1 = 2$, $\ell_2 = 4$ $\ell_3 = -2$. For visual clarity, both traces are normalized to the value of the maximum sample. In the case where $\ell_1 = 0$, $\ell_2 = 0$ and $\ell_3 = 0$ the echo is dominated by all of the other pulses. When the programming and probe fields are in OAM modes, they are mode filtered by the single mode fiber and are attenuated between 17 dB and 29 dB leaving the SPE as the dominant field.}
\end{figure}

\paragraph{}
\textit{Introduction}---The multimodal optical storage and signal processing capacity of cryogenic rare-earth ion (REI) doped crystalline solids makes them attractive candidates for quantum memory and quantum signal processing platforms \cite{thiel_rare-earth-doped_2011,ortu_multimode_2022}. In particular, the atomic frequency comb (AFC) protocol has been widely explored as a route to a quantum memory system that could be deployed as a functional quantum repeater\cite{feldmann_cavity-enhanced_2025,masuko_frequency-multiplexed_2024}. REI systems have also been demonstrated in a variety of classical signal processing applications, ranging from spectrum analysis\cite{berger_rf_2016,krishna_mohan_ultra-wideband_2007} and geolocation of RF signals\cite{barber_angle_2010} to high bandwidth real time correlation \cite{babbitt_coherent_1994,barber_spatial-spectral_2017} and microwave to optical conversion\cite{chaneliere_opto-rf_2024}. Demonstrations of complex spatial modes used in stimulated photon echo experiments were conducted as early as the 1980s \cite{saari_picosecond_1986}, but applications of structured light in REI systems has not been fully explored. Despite a strong theoretical understanding  of the multimodal properties of these materials, there have been few experimental demonstrations utilizing orbital angular momentum (OAM) as a degree of freedom.
\paragraph{}
Since their inception, optical OAM beams have been explored for a variety of applications in free-space communications. Optical OAM provides a convenient basis for spatial division multiplexing (SDM) in free space optical communication \cite{wang_terabit_2012,lei_massive_2015} while in the realm of quantum communications, OAM is a high dimensional degree of freedom that can be used for entanglement\cite{doi:10.1098/rsta.2015.0442,mair_entanglement_2001} and hyper-entanglement\cite{PhysRevA.105.062422}. Additionally, there is active research in the application of OAM carrying beams to minimize wander and scintillation, utilizing their possible self regenerative properties when exposed to turbulent conditions\cite{PhysRevLett.94.153901,BOUCHAL2002155,Gbur:08}. All of these properties could be exploited when combined with the signal processing capabilities of REI systems, and in this work, phase matching condition for OAM in a stimulated photon echo (SPE) process is demonstrated experimentally. 
\paragraph{}
\textit{Stimulated Photon Echos}---Consider a a pair of broadband optical pulses, the first $E_1(\bm{r},t)e^{i(\bm{k}_1 \cdot \bm{r}-\omega_0t)}$ and $E_2(\bm{r},t+\tau_{21})e^{i(\bm{k}_2 \cdot \bm{r}-\omega_0(t-\tau_{21}))}$ which is delayed from the first pulse by $\tau_{21}$ which is less than the material coherence lifetime ($t_2$). When these pulses illuminate an REI material they will burn an interference pattern into the inhomogeneously broadened absorption profile, with a sinusoidal alteration of the form presented in Eq. \ref{eq:grating}.
\begin{equation}\label{eq:grating}
    \Delta\alpha(\bm{r},\omega)\propto E_1^*(\bm{r},\omega)E_2(\bm{r},\omega)e^{i((\bm{k}_2-\bm{k}_1)\cdot \bm{r}-\omega\tau_{21})}+c.c.
\end{equation}
This pattern, referred to as a spatial-spectral grating, can be used to spatially and temporally diffract another pulse. Consider a third pulse, $E_3(\bm{r},t+\tau_{32}+\tau_{21})e^{i(\bm{k}_3 \cdot \bm{r}-\omega_0(t-\tau_{32}-\tau_{21}))}$, that interacts with the grating at a time $\tau_{32}$ after the second pulse. A fourth pulse will be generated as a consequence of interaction at a time $\tau_{21}$ after this third pulse. This fourth pulse is known as an SPE, with the general form in frequency domain given in Eq. \ref{eq:echo_form}.
\begin{eqnarray}\label{eq:echo_form}
    E_{echo}(\bm{r},\omega)\propto  E_1^*(\bm{r},\omega)E_2(\bm{r},\omega)E_3(\bm{r},\omega)\\ 
    \times  e^{i((\bm{k}_2+\bm{k}_3-\bm{k}_1) \cdot \bm{r}-\omega_0(t-\tau_{32}-2\tau_{21}))} \nonumber
\end{eqnarray}
\paragraph{}
\textit{Optical Orbital Angular Momentum}--- OAM beams have helical wavefronts specified by an integer parameter $\ell$, often referred to as a topological charge. This is achieved by applying an azimuthally varying spatial phase shift across the wavefront $\phi(\rho,\theta)=\ell\theta$. 
This topological charge governs the optical OAM of each photon in an OAM beam, which is given by $\bm{L}=\hbar\ell\hat{k}$ \cite{yao_orbital_2011}. 
\paragraph{}
\textit{Orbital Angular Momentum Phase Matching in Stimulated Photon Echoes}---In the SPE process, the linear momentum per photon of the echo is governed by the phase matching condition $\bm{k}_{echo}=\bm{k}_2+\bm{k}_3-\bm{k}_1 \to \bm{P}_{echo}=\bm{P}_2+\bm{P}_3-\bm{P}_1$. This phase matching condition applies at every point across the face of the beam. By taking a cross-product with position and averaging over the beam front to calculate the OAM per photon, it can be inferred that the OAM must follow a similar phase matching condition, which is explicitly given in Eq. \ref{eq:L_phase_matching}.
\begin{equation}\label{eq:L_phase_matching}
\bm{L}_{echo}=\bm{L}_2+\bm{L}_3-\bm{L}_1
\end{equation}
Alternatively, OAM can be accounted for in an SPE process by assuming a sequence of three coaxial pulses with the same wave-vector  $E_1,E_2$ and $ E_3$ with topological charges $\ell_1,$ $\ell_2$ and $\ell_3$ respectively. The first two pulses will program a spectral grating with an azimuthal phase dependence $\phi_{grating}(\theta)=(\ell_2-\ell_1)\theta$. The spectral grating will then impart OAM on the SPE. The topological charge of the SPE is given by Eq. \ref{eq:echo_topological_charge}.
\begin{equation}\label{eq:echo_topological_charge}
    \ell_{echo}=\ell_{2}+\ell_3-\ell_1
\end{equation}
\paragraph{}
\textit{Generation of multiplexed OAM states}---To demonstrate the conservation of topological charge in the stimulated echo, a Dammann vortex grating (DVG) is displayed on an SLM (Holoeye) such that each input field carries a unique charge \cite{lei_massive_2015} that would result in the generation of an SPE with no topological charge. The DVG consists of two orthogonal binary forked grating phasors multiplied together modulo 2$\pi$\cite{moreno_encoding_2010}, such that there are two unique charges associated with the combined grating, $\ell_{x}$ and $\ell_{y}$. This results in an incoming beam imaged at the center of the grating diffracting into four primary modes, each with a different topological charge produced as a sum or difference of the individual binary forked gratings. When three beams in a traditional box geometry\cite{barber_spatial-spectral_2017} are imaged onto the DVG, there will be a single point in which a diffracted order from each beam will spatially overlap, and a pictorial representation of this process can be seen in Fig. \ref{fig:DVG_Schematic}. This results in a single collinear beam that minimizes the complexity of optical design when interfacing with the rare earth ion system.
\paragraph{}
In this letter, we demonstrate the ability to generate SPEs with unique topological charge based on the temporal phase matching conditions that occur within a sample of Tm\(^{3+}\):YAG (Teledyne FLIR: Scientific Materials), cooled to 3.4 K using a CA-100 cryostation (Montana Instruments). We also demonstrate multiple degenerate cases in which the result of Eq. \ref{eq:echo_topological_charge} is identical for various input fields. Lastly, we utilize spatial filtering to show how this technique can be used to increase the signal-to-background ratio of the SPE with respect to the input pulse sequence.
\paragraph{}
\textit{Detection of Stimulated Echos}---To access the inhomogeneously broadened absorption spectra of the material, an L-band fiber laser (NKT photonics) is amplified and frequency doubled with a periodically poled lithium niobate waveguide (AdVR). To program the material with the required optical sequence, the near infrared laser is split utilizing a fiber coupled beamsplitter array (ThorLabs) before each individual beam enters a fiber coupled acousto-optic modulator (Brimrose) driven by an RF pulse generator (BNC) to create the required pulse sequences to produce an SPE. These pulses transmit into a free space optical system as illustrated in Fig. \ref{fig:OAM_Echo_Schematic}. A DVG with the parameters $\ell_{x}=1$ \& $\ell_{y}=3$ is generated on the SLM such that the SPE would have $\ell=0$ based on Eq. \ref{eq:echo_topological_charge} when combined with the coaxial geometry generated in Fig. \ref{fig:DVG_Schematic}. The results of this measurement can be seen in Fig.  \ref{fig:OAM_Images}, which shows the intensity of the programming pulses as well as the stimulated echo before and after an astigmatic transform. The topological charge is deduced from the astigmatically transmitted beam by counting null fringes \cite{alperin_quantitative_2016}.
\paragraph{}
\textit{Spatial Filtering for Stimulated Echo Recovery}---To demonstrate an immediate advantage over traditional REI based signal processing geometries, the cylindrical lens and CMOS camera shown in Fig. \ref{fig:OAM_Echo_Schematic} are replaced with a fiber coupled variable gain optical detector. This configuration allows for the capture of time domain traces of both the programming sequence as well as SPE, where it becomes possible to compare echo strength and optical extinction ratio in the traditional gaussian programming conditions vs. the colinear OAM programming condition. Given the optical mode matching conditions of the single mode fiber, fields carrying topological charge $\ell$ inefficiently couple into the fiber, while the stimulated echo with no charge will have maximal coupling efficiency. Fig. \ref{fig:time_domain_echoes} presents the normalized traces of both the gaussian programming condition as well as the OAM programming condition, and Table \ref{tbl:Spatial_Filtering_Results} presents the improvements in the echo to programming pulse ratios as well as extinction ratios for each optical field detected. 
\begin{table}[h]
\caption{Spatial Filtering Efficiency}
\begin{ruledtabular}
\begin{tabular}{cccc}
Pulse \footnotemark[1]              & \multicolumn{2}{c}{Echo to Pulse Ratio} & Extinction Ratio\footnotemark[2] \\ \hline
                   & Gaussian            & OAM               &                  \\
                   & Sequence            & sequence          &                  \\ \cline{2-3}
Pulse 1 ($\ell=2$) & -16.14 dB            & 4.20 dB            & -21.67 dB         \\
Pulse 2 ($\ell=4$) & -15.83 dB            & 11.96 dB           & -29.12 dB         \\
Pulse 3 ($\ell=2$) & -9.43 dB             & 6.48 dB            & -17.24 dB         \\
SPE ($\ell=0$)     &                     &                   & -1.33 dB         
\end{tabular}
\end{ruledtabular}
\footnotetext[1]{The topological charge of each pulse is specified for the OAM sequence case.}
\footnotetext[2]{Extinction ratio for each pulse is computed as a ratio of total pulse energy between the gaussian and OAM sequence.}
\label{tbl:Spatial_Filtering_Results}
\end{table}
\paragraph{}
\textit{Discussion and Conclusions}--- The demonstrated OAM phase matching condition allows for modal separation of programming fields and stimulated echoes. Other techniques to isolate SPEs require multiple paths\cite{wolfe_demonstration_2019} or complex geometries\cite{barber_spatial-spectral_2017}, but using optical OAM requires only a single coaxially multiplexed beam. Additional fields can be readily multiplexed allowing for more unique programming conditions. In the classical regime this corresponds to multiple programming templates in an optical correlator, while in the quantum regime this has applications in generating selective atomic frequency combs for quantum routing. 
\paragraph{}
The demonstration in Fig. \ref{fig:OAM_Images} details a case where $\ell_{echo}=0$, but it cannot be understated that the novel phase matching condition for topological charge readily admits solutions with a unique non-zero OAM. This has immediate applications in state preparation for quantum transport in free-space telecommunication systems, as well as the generation of hyper-entangled optical qubits. By exploiting this phase matching condition in a traditional AFC protocol to ensure recalled qubits have a non-zero topological charge, unwanted photons including those from scattered light or fluorescence (which is incoherent and therefore will not retain OAM) could be suppressed. This requires no change to the spectral and temporal programming conditions and only requires passive spatial phase modulation. 
\paragraph{}
To conclude, we experimentally demonstrate a novel phase matching condition for OAM in SPEs. A pulse sequence exploiting this phase matching condition was used to generate SPEs with a unique topological charge. By enforcing conditions where $\ell_{echo}=0$ we demonstrate the isolation of the echo pulse through modal filtering resulting in a signal-to-background ratio of 6.4 dB when comparing the echo to pulse 3. This compares favorably with respect to interferometric echo isolation, which has been demonstrated to provide 6 dB signal-to-background \cite{wolfe_demonstration_2019}, but requires active phase stabilization unlike the method presented in this letter.

\textit{Acknowledgments}---Blender assets courtesy of Ryo Mizuta Graphics.
This project was sponsored in part by the Air Force Research Laboratory under Agreement Number FA8750-24-1-1019 \& FA8750-23-20500. The U.S. Government is authorized to reproduce and distribute reprints for government purposes notwithstanding any copyright notation thereon. Any opinions, findings, and conclusions or recommendations expressed in this material are those of the authors and do not necessarily reflect the view of the funder.
\bibliography{apssamp}
\section{End Matter}
\begin{figure*}[h]
\includegraphics[width=.8\textwidth]{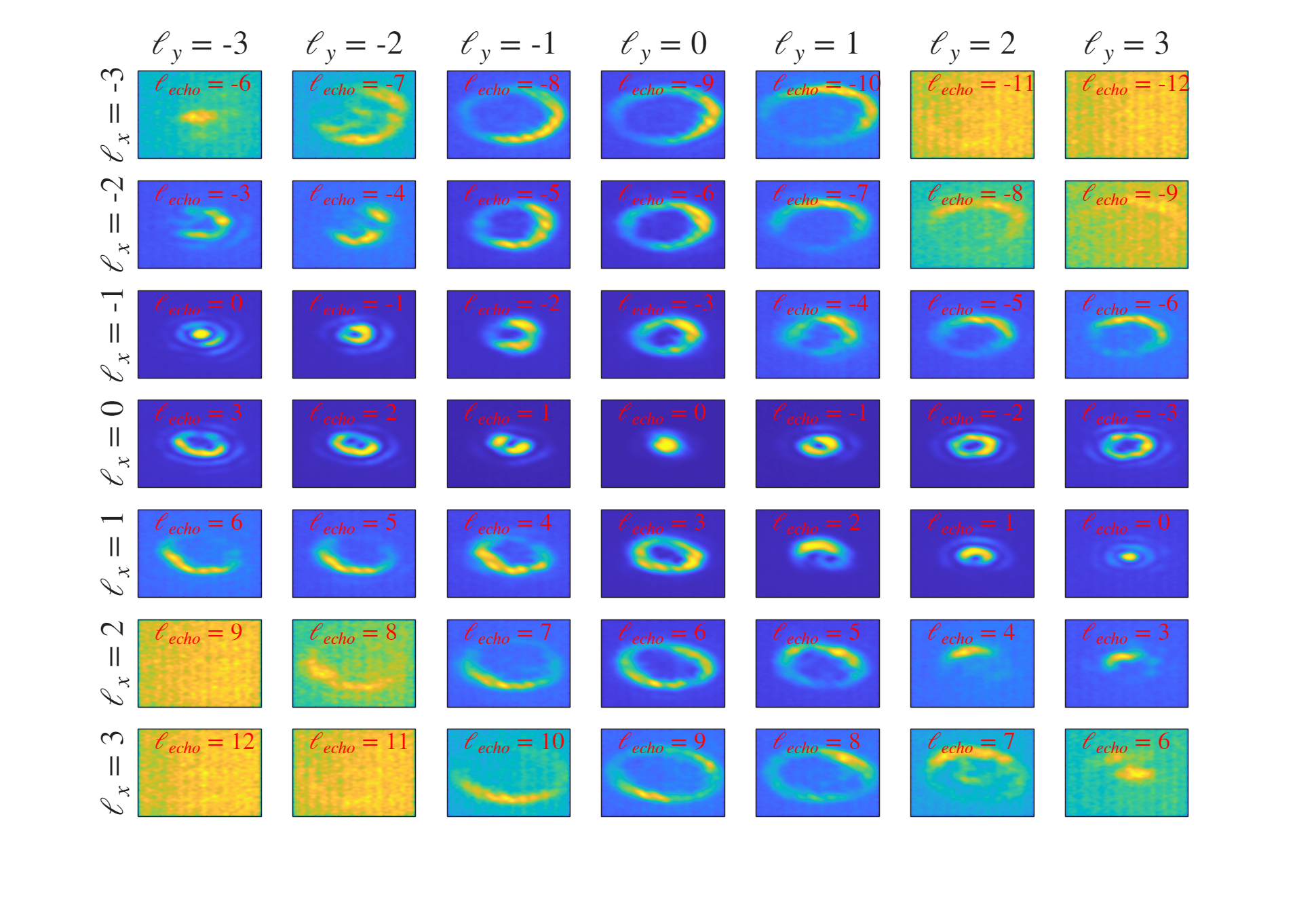}
\caption{\label{fig:image_array} An array of images captured using the CMOS camera showing the intensity distribution of the SPE as a function of DVG parameters. Images are indexed according to the x and y topological charge values of the grating used to multiplex input pulses. Each image is also marked with the topological charge of the SPE, according to the parameters of the DVG.}
\end{figure*}

\begin{figure*}
\includegraphics[width=.6\textwidth]{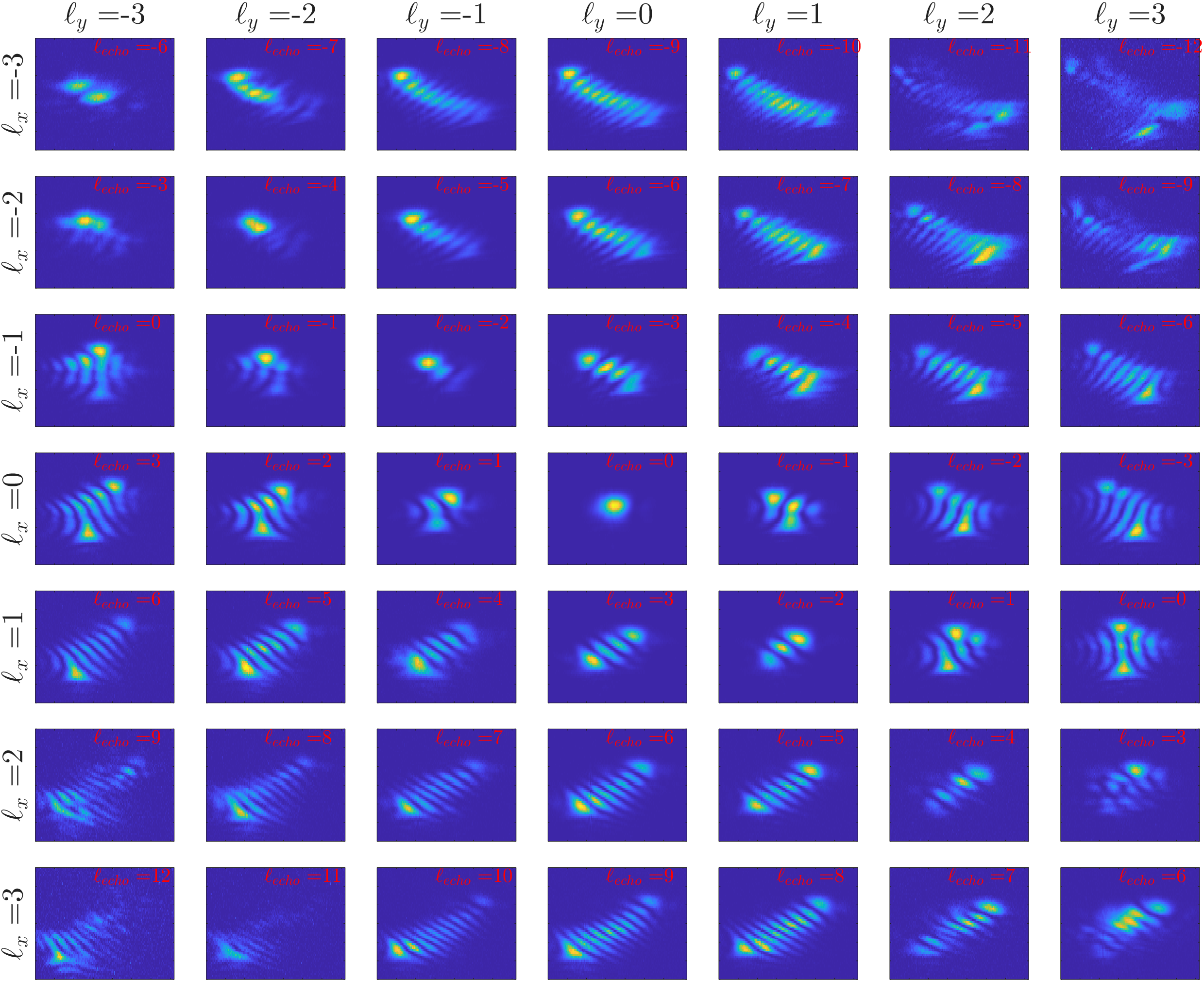}
\caption{\label{fig:Astigmatic_Image_Array} An array of images captured using the CMOS camera after performing an astigmatic transform via cylindrical lens. Null fringe counting is used to verify that the topological charge matches the expected charge as dictated by the OAM phase matching condition discussed in this work.}
\end{figure*}

\textit{Appendix: SPE generation with multiple DVG inputs}---To verify the successful generation of SPEs that carry a topological charge greater than zero, multiple DVGs were deployed across a range spanning $-3\leq\ell_{x},\ell_{y}\leq3$. Following the phase matching condition from Eq. \ref{eq:echo_topological_charge}, the topological charge of the SPE was bounded between $-12\leq\ell_{echo}\leq12$, and images of each echo intensity were recorded and are shown in Fig. \ref{fig:image_array}. 
It is worth noting that in the ideal case, the presented matrix would result in equal intensity echos for the positive and negative topological charges, but fluctuations in optical power for the three input beams reduce the SPE efficiency. After imaging the echo, the cylindrical lens was reintroduced into the beam path, and captures of the astigmatic transform for each echo were recorded, which are shown in Fig. \ref{fig:Astigmatic_Image_Array}
Most of the astigmatic transforms can be readily interpreted to estimate topological charge for each field, however for some of the higher order cases, where the pulses carry higher topological charge, the data is slightly harder to interpret. This indicates that there may be some clipping of higher order fields OAM modes on parts of the experimental apparatus. Given the phase matching condition and the configuration of the DVG, the expected topological charge of the echo is $\ell_{echo}=3\ell_x-\ell_y$.
Aside from issues with higher order OAM beams, the echoes exhibit the expected fringe pattern given the expected topological charge  OAM phase matching conditions. 
\end{document}